\documentclass[reprint,%superscriptaddress,amsmath,amssymb,aps,
aps,twocolumn]{revtex4}
\usepackage{amssymb}
\usepackage{amsmath}
\usepackage{graphicx}% Include figure files
\usepackage{bm}% bold math
\newcommand{\beq}{\begin{equation}}
\newcommand{\eeq}{\end{equation}}

\newcommand{\bee}{\begin{eqnarray}}
\newcommand{\eee}{\end{eqnarray}}
\newcommand{\ba}{\begin{array}}
\newcommand{\ea}{\end{array}}
\newcommand{\bc}{\begin{center}}
\newcommand{\ec}{\end{center}}
\newcommand{\bi}{\begin{itemize}}
\newcommand{\ei}{\end{itemize}}

\usepackage{color}

\begin{document}

\title{Periodic and quasiperiodic motions of many particles falling in a viscous fluid}

\author{Marta Gruca} 
\author{Marek Bukowicki}
\author{Maria L. Ekiel-Je\.zewska}
\email{mekiel@ippt.pan.pl}
\thanks{Corresponding author}
\affiliation{Institute of Fundamental Technological Research, Polish Academy of Sciences, Pawi\'nskiego 5b, 02-106 Warsaw, Poland}

\date{\today}

\begin{abstract}
Dynamics of regular clusters of many non-touching particles falling under gravity in a viscous fluid at low Reynolds number are analysed within the point-particle model.  
Evolution of two families of particle configurations is determined: 2 or 4 regular horizontal polygons (called `rings') centered above or below each other. Two rings fall together and periodically oscillate. Four rings usually separate from each other with chaotic scattering. For hundreds of thousands of initial configurations, a map of the cluster lifetime is evaluated, where the long-lasting clusters are centered around periodic solutions for the relative motions, and surrounded by regions of the chaotic scattering, %. The results illustrate that for systems of many sedimenting particles, there exist periodic relative motions which are important for the dynamics for a wide range of close by initial configurations, with quasi-periodic oscillations which keep the particles together for very long times, and chaotic scattering, 
in a similar way as it was observed by Janosi et al. (1997) for three particles only. These findings suggest to consider the existence of periodic orbits as a possible physical mechanism of the existence of unstable clusters of particles falling under gravity in a viscous fluid.
\end{abstract}

\maketitle
\section{Introduction}\label{Intro}
In various contexts of the low-Reynolds-number fluid mechanics, periodic motions have
turned out to be essential for the dynamics of particle
systems. Jeffery orbits of elongated particles or pairs of spherical particles in shear flow provide a classical example \cite{Jeffery}. 
There has been also a lot of interest in oscillatory motions of 
several particles under external force fields. A modern example is a system of several particles kept by external forces inside a toroidal optical trap.
Periodic oscillations, hydrodynamic pairing of particles, limit cycles, and transition to chaos have been found in experiments and theoretical models \cite{Stark,Stark2,Roichman1,Roichman2,Sassa}.

For many decades, periodic settling of a small number of
particles under gravity has been extensively studied experimentally \cite{Kaye,Jayaweera,Koglin1,Koglin2,Koglin3,Koglin4,Alabrudzinski,Nowakowski} and 
theoretically  \cite{Hocking,Brady,Caflisch,Tory4,Tory,Golubitsky,Tory2,Lim,Snook,Szymczak,Bargiel,Tory3}, based on the 
Stokes equations for the fluid flow. A new perspective for the role of such benchmark solutions 
was opened by 
%Recently, .....%....... 
%Back to sedimentation. A lot of papers on periodic motion 
%
Janosi et al. \cite{Janosi} who showed that
three particles settling under gravity in a viscous fluid
in a vertical plane perform chaotic scattering, and 
related this behaviour to existence of an unstable periodic solution (which, however, was not directly found). For
a wide range of arbitrary random initial configurations,
the particle stay together if they are sufficiently close
to this periodic solution; the closer they are, the longer
they stay together  \cite{Janosi}. Later, it was shown that three spheres 
exhibit a similar chaotic scattering \cite{EJ}, and for spherical particles, the periodic solutions related to chaotic scattering have been explicitly found 
\cite{Ekiel-Jezewska2}. 

It is important to check if a similar behaviour can be also observed in case of clusters made of a very large number of particles: is chaotic scattering of many particles at non-regular configurations also observed, and is it coupled to periodic oscillations of regular arrays? Is the destabilization pattern of a suspension drop \cite{Batchelor,Mylyk} related to the existence of periodic solutions for relative motions of the particles in certain regular configurations, as suggested in Ref.~\cite{Ekiel-Jezewska}? To address these open challenging problems, the first step %to gain some understanding 
is to find and analyze examples of periodic oscillations of a large number of particles settling under gravity in a viscous fluid.

In Ref.~\cite{Brady}, periodic oscillations of a cube with the walls parallel and perpendicular to gravity were reported, and it was supposed that other regular polyhedrons may display similar behaviour. However, it has turned out that the periodic orbits of 
the cube~\cite{Brady} generalize to another family of periodic oscillations of an arbitrary large (even) number of particles: the particles arranged in two mirror regular horizontal polygons, one above the other~\cite{Ekiel-Jezewska}. %Obviously, the cube is an example which belongs to this family. %This family is a non-straightforward generalization of 
For a similar geometry of arrays of rods (and other non-spherical particles), centered at vertices of a regular horizontal polygon, periodic solutions have been also found, both experimentally and theoretically~\cite{Jung}.

In this paper, we start from generalizing the results of \cite{Ekiel-Jezewska}, where periodic solutions were explicitly found for 
a moderate number of particles centered at vertices of two regular horizontal polygons (`2 rings'). In this work, we explicitly demonstrate that 2 rings made of a very large number %thousands
of particles indeed fall oscillating periodically. Next, we modify the initial particle positions to obtain `less regular' configuration, or in other words - we desynchronize the motion of the particles, and investigate if periodic solutions still can be found, and if they are related
to the existence of clusters with long lifetimes and chaotic scattering of particles in configurations which are close to the periodic orbits.

The modification of the initial conditions is based on the idea of Ref. \cite{Ekiel-Jezewska}, where the 2-rings configurations were desynchronized by moving every second particle from each ring to the position it would have after one fourth of the period. Such a perturbation resulted in the initial configuration of four regular horizontal polygons (`4 rings'), for which  long-lasting quasiperiodic relative 
motions of particles were reported \cite{Ekiel-Jezewska}. The question is if there exist some periodic solutions for an initial configuration close by. Therefore, in this work we 
investigate in details the dynamics of 4 rings for a wide range of 
the initial conditions and different numbers of particles. We evaluate a map of the cluster lifetimes, search for periodic solutions and chaotic scattering.

\section{System and its theoretical description}\label{sec:model}

We investigate dynamics of regular groups of 
many point-particles settling under identical gravitational forces $\bm{G}$ in a fluid of viscosity $\eta$ at the low-Reynolds-number. 
The fluid velocity $\bm{v}$ and pressure $p$ satisfy the Stokes equations, see e.g. \cite{KimKarrila},
\begin{eqnarray}
\label{stokes_eq}
\eta \nabla^2 \bm{v}(\bm{r}) - \nabla p(\bm{r}) &=& -\sum_{i = 1}^{M} \bm{G}\,\delta (\bm{r} - \tilde{\bm{r}}_i),\\ 
\nabla \cdot \bm{v(r)} &=& 0.
\end{eqnarray}
where $M$ is the number of particles, $\tilde{\bm{r}}_i$ denotes the position of the particle with label $i$, and the $z$-axis is chosen
along gravity, with $\bm{G}=-G \hat{\bm{z}}$ where $G>0$ and $\hat{\bm{z}}$ is the unit vector along the $z$-axis. The equations are written in the laboratory frame of reference.

The fluid velocity $\bm{v}(\bm{r})$ and pressure $p(\bm{r})$ %in the laboratory frame of reference 
are given by
\begin{eqnarray}
 \bm{v}(\bm{r}) &=& \sum_{i=1}^M \bm{T}(\bm{r} - \tilde{\bm{r}}_i)\cdot \bm{G}, \\
 p(\bm{r}) &=& \sum_{i=1}^M \bm{P}(\bm{r}-\tilde{\bm{r}}_i)\cdot \bm{G},
\end{eqnarray}
with the Green tensors,
\begin{eqnarray}
 \bm{T}(\bm{R}) =  \frac{1}{8\pi\eta R} \left(\bm{I} + \frac{\bm{R} \otimes \bm{R}}{R^2}\right), \\
 \bm{P}(\bm{R}) = \frac{\bm{R}}{4\pi R^3},
\end{eqnarray}
where $R=|\bm{R}|$.
%with $\bm{R} = \mathbf{r - r_i}$.

%With this choice, 
The equations of motion of the particles are given by
\begin{equation}
	\frac{d\tilde{\bm{r}}_i}{dt} =  \sum_{j\neq i}^{M} \bm{T}(\tilde{\bm{r}}_{ij}) \cdot \bm{G}
	%\hat{\bm{z}} 
	\label{dyn} + \mathbf{u}_0
\end{equation}
where  $\tilde{\bm{r}}_{ij} = \tilde{\bm{r}}_i - \tilde{\bm{r}}_j$, 
and $\mathbf{u}_0 = {\bm{G}}/{(6\pi\eta a)}$ is the Stokes velocity of the isolated particle. 
While solving Eqs.~\eqref{dyn}, it is convenient to follow \cite{Ekiel-Jezewska} and 
choose the inertial frame of reference moving with the velocity $\mathbf{u}_0$. The benefit is that in this frame the dynamics is independent of 
the additional length scale $a$. The transformation back to the laboratory frame is straightforward, but not needed in this work, where we analyze the dynamics in the centre-of-mass frame of reference. 

From now on, we use dimensionless variables, based on an initial size of the group $d$ as the length unit, and ${G}/(8 \pi \eta d)$ as the velocity 
unit. 
Therefore, ${8 \pi \eta d^2}/G$ is the time unit. From now on, $\bm{r}_i=\tilde{\bm{r}}_i/d$ denotes the dimensionless position of a particle $i$.

\begin{figure*}
\includegraphics[width=11cm]{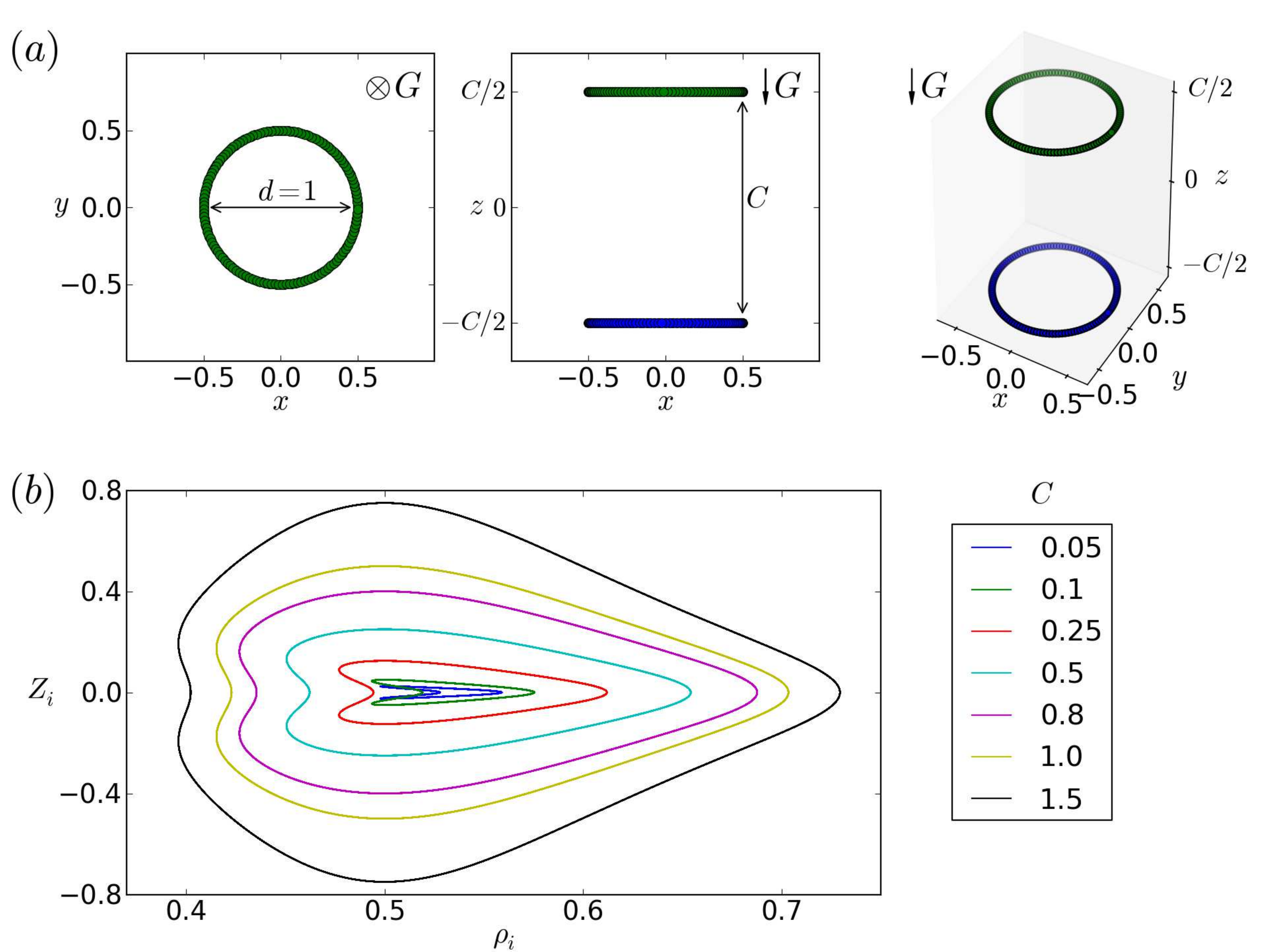}
%\label{fig:schem_2rings}
\caption{(Color online) The system of $M$ particles which form 2 rings. (a) $M\!=\!256$.  The initial configuration as in Eq.~\eqref{2in}. 
(b) $M\!=\!20000$. In the centre-of-mass frame, the trajectory of each particle has the same shape (not drawn to scale). Different shapes 
correspond to the indicated values of $C$.}
\label{fig:traj_2rings_20000par}
\end{figure*}

Initially, the particles are placed at vertexes of $K$
horizontal regular polygons (called 'rings') which are separated from each other vertically and centered one above (or below) the other. 
The diameter $d$ of the top ring is taken as the length unit. 
The number $N$ of particles in every ring is the same. Therefore, the total number of particles in the system is equal to $M=KN$.  Because of the system symmetry, we use the cylindrical coordinate system, in which $\bm{r}_i$ is represented as $\bm{r}_i=(\rho_i, \phi_i, z_i)$.

Searching for periodic solutions, we take into account that numerical non-symmetrical perturbations can destroy periodic unstable solutions after times smaller than the period, as observed in Ref. \cite{Ekiel-Jezewska}. 
Therefore, 
we symmetrize the dynamics: 
we force the azimuthal components of the particle velocities to vanish, with $\phi_i\!=\!\text{const}(t)$ for $i\!=\!1,...,M$, and the polygons to remain regular for all times. We use the parametrization $i\!=\!K(n-1)+k$, with $n\!=\!1,...,N$ and $k\!=\!1,...,K$. Then,
\bee
\phi_{K(n-1)+k}=\frac{2\pi(n-1)}{N},
\eee
and  the radial and vertical  coordinates of the particles 
from the same polygon $k$ are the same, 
\bee
\rho_{K(n-1)+k}\!&=&\!\rho_k,\\ 
z_{K(n-1)+k}\!&=&\!z_k,
\eee
for $n\!=\!1,...,N$ and $k\!=\!1,...,K$. 
The system of $3M$ equations \eqref{dyn} is reduced to the system of $2K$ equations for $\rho_l$ and  $z_l$, with $l=1,...,K$, which in the frame of reference moving with the velocity $\mathbf{u}_0$ can be explicitly written as,
\begin{eqnarray}
 \frac{d\rho_l}{dt}\! &=&\!\! - \sum_{k = 1}^{K} \sum_{n = 1}^{N} \frac{(z_k -
z_l)\left[\rho_k \cos\left(\frac{2\pi(n-1)}{N}\right) - \rho_l\right]}{R_{lkn}^3}, \label{EOM}\\
 \frac{dz_l}{dt} \!&=& \!\!- \sum_{k = 1}^K \sum_{n = 1}^N \left( \frac{1}{R_{lkn}}
+ \frac{(z_k - z_l)^2}{R_{lkn}^3} \right), \\
 \! \! R_{kln}^2\!\! &=&\! (z_k - z_l)^2 + \rho_l^2 + \rho_k^2 - 2\rho_l \rho_k
\cos\left(\!\frac{2\pi (n-1)}{N}\right)\!.\;\nonumber\\\label{EOM3}
\end{eqnarray}

%The equations \eqref{dyn} are solved numerically. 
The numerical integration of Eqs. \eqref{EOM}-\eqref{EOM3} is based on two methods:
the fourth-order adaptive Runge-Kutta for non-stiff problems and backward differentiation formula 
(BDF) for stiff problems. We solve a system of ordinary differential equations using lsoda from the FORTRAN library odepack. Initially we 
start with non-stiff solver. Depending on number of time steps and accuracy the solver switches between method appropriate for stiff or 
non-stiff problem. This procedure is a standard and commonly used method for solving differential equations. This solver is sufficiently 
stable~\cite{Numerical}. The numerical calculations were performed with double precision and the error per each time step is not 
greater than $10^{-12}$.

 We are interested in relative motion of the particles; therefore, we will trace their positions in the centre-of-mass frame of reference, located at the symmetry axis, $\bm{r}_{CM}=(0, 0, z_{CM})$. In this frame, $\rho_i$ and $\phi_i$ are the same as in the laboratory frame, and particle vertical coordinates are $Z_i=z_i-z_{CM}$.

We perform
simulations for two families of the systems: with two and four rings. The specific initial configurations and their evolution 
will be discussed in the next sections.

\section{Dynamics of 2 rings}\label{sec:2_rings}
We first consider a system of $M=2N$ particles which are grouped in two horizontal rings, each containing $N$ equally spaced particles. 
Initially, the rings are identical and placed one exactly above the other. 
The diameter of each ring 
is equal to $1$ and the initial distance $C$ between the rings 
is a parameter 
in our
simulations, as shown in figure \ref{fig:traj_2rings_20000par}$(a)$. The initial positions $\bm{r}_i=(\rho_i, \phi_i, z_i)$ of the particles $i=1,...,M$  are 
\bee
\bm{r}_{2n-1}&=&\left(\frac{1}{2},\frac{2\pi (n-1)}{N},\frac{C}{2}\right),\\ 
\bm{r}_{2n}&=&\left(\frac{1}{2},\frac{2\pi (n-1)}{N},-\frac{C}{2}\right),\label{2in}
\eee
with $n=1,...,N$. 
We conducted around 100 simulations for $M=256$ and $20 000$ particles and 
different initial values of $C \in [0.05,2.5]$. 
For these values of the parameters, 
we observe periodic motion of the particles 
during the whole simulation time $t=5000$ (what corresponds to 300-2700 periods), %as shown in movie \ref{fig:traj_2rings_20000par}.
see our movie \ref{fig:traj_2rings_20000par} in \cite{movies}. 

In the centre-of-mass frame,   
two particles 
with the same
angular coordinates 
move along the same trajectory. Its shape is 
the same for all the pairs, and it depends on $C$ as shown in figure \ref{fig:traj_2rings_20000par}$(b)$ for $M\!=\!20~000$. We found similar
families of shapes for $M\!=\!256$, analogical to those reported by \cite{Ekiel-Jezewska} for $M\!=\!16$, and resembling periodic solutions 
for rigid rods, discussed by 
\cite{Jung} and \cite{Bukowicki}.
\par

\begin{figure*}
\includegraphics[width=12cm]{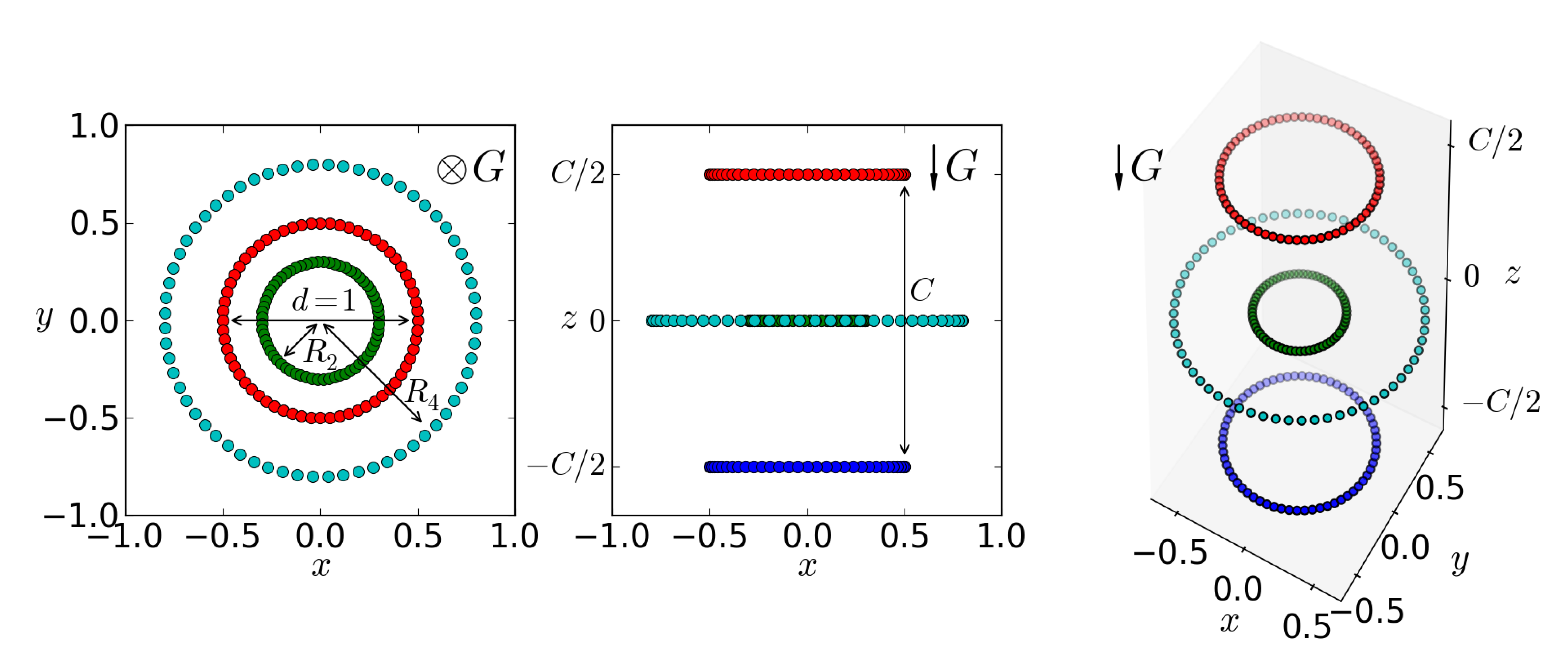}
\caption{(Color online) Initially, $M$ particles form 4 rings, as  specified in Eq.~\eqref{4in}. Here, $M\!=\!256$.}\label{fig:schem_4rings} 
\end{figure*}
\begin{figure*}
\includegraphics[width=\textwidth]{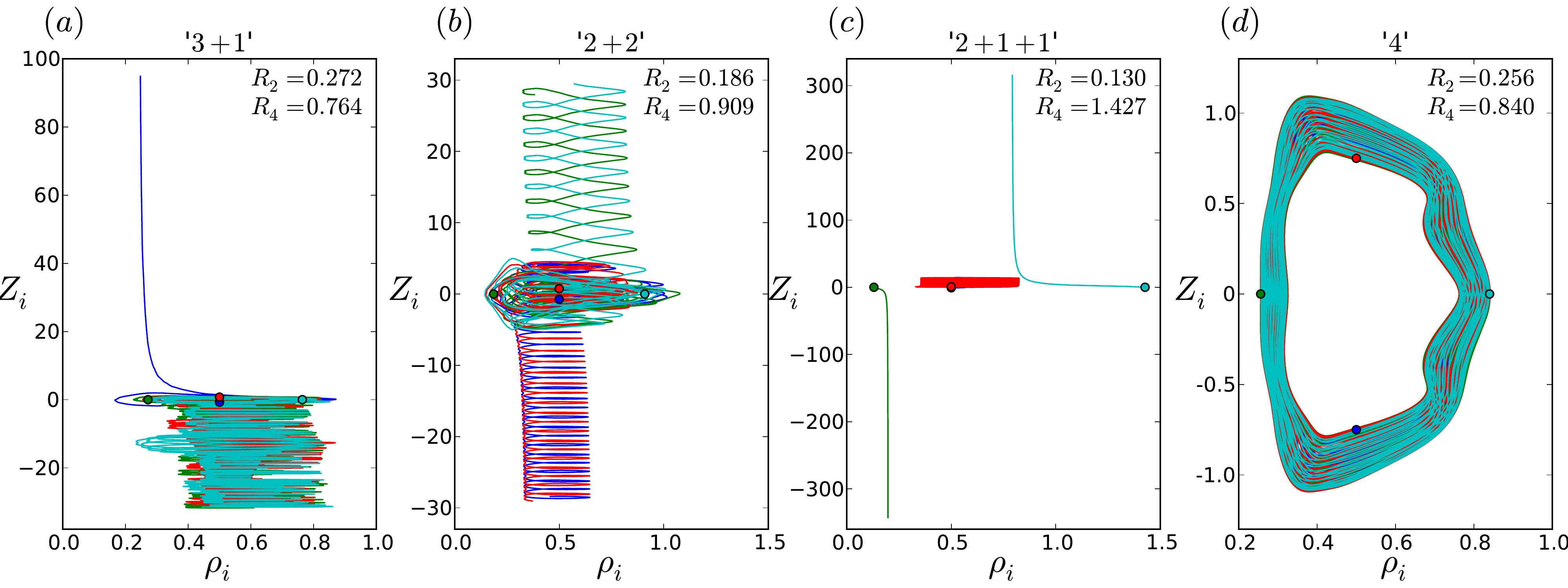}
\vspace{-0.5cm}
\caption{(Color online) Evolution of $M\!\!=\!\!256$ particles which form 4 rings with $C \!=\! 1.5$ and $R_2$, $R_4$ as indicated. Trajectories of four particles (each from a different ring) in the centre-of-mass reference frame until time $t\!=\!500$. The initial positions  %of the particles 
are 
marked by dots. (a) `3+1' type of decay. One of the rings separates from the cluster. 
(b) `2+2' 
type of decay. The group breaks up into two pairs of oscillating rings - one of them falls faster than the other.  (c) `2+1+1' type of decay. The cluster
breaks up into one pair of oscillating rings and two single rings. (d) `4': lack of decay. All rings oscillate. 
}\label{fig:tra_4} 
\end{figure*}

\section{Dynamics of 4 rings}\label{sec:4_rings}
\subsection{Initial configurations}

In the second system, $M=4N$ particles are grouped in four horizontal rings, each consisting of $N$ particles. Initially, the rings labeled $1$ and $3$ are
 placed 
at $z = {C}/{2}$ and $-C/2$, respectively, and the rings labeled $2$ and $4$: at $z = 0$, as shown in figure \ref{fig:schem_4rings}$(a)$.  
The radii of the rings $1,3,2$ and $4$ are equal to $1/2$, $1/2$, $R_2$,  and $R_4$, respectively, with $R_4 > R_2$.  The angular coordinates of the particles from ring $1$ 
 and ring $3$ 
are the same. The other rings 
are rotated by ${\pi}/{N}$ around the symmetry axis.
Summarizing, the initial cylindrical coordinates of the particles are,
\begin{eqnarray}
\bm{r}_{4n-3}\! &=& \!\left(\frac{1}{2}, \frac{2\pi (n-1)}{N},\frac{C}{2}\right),\\
\bm{r}_{4n-2}\! &=& \!\left(R_2, \frac{2\pi (n-\frac{1}{2})}{N},0\right), %\hspace{0.5cm}
\label{4in_a}
\end{eqnarray}
\begin{eqnarray}
\bm{r}_{4n-1}\! &=&\! \left(\frac{1}{2}, \frac{2\pi (n-1)}{N},-\frac{C}{2}\right),\\
\bm{r}_{4n}\! &=&\! \left(R_4,\, \frac{2\pi (n-\frac{1}{2})}{N},0\right), %\\
\label{4in}
\end{eqnarray}
where $n\!=\!1,...,N$ and 
$R_2$ and $R_4$ are the parameters of the simulation.

We performed around 400~000 simulations for clusters made of different numbers of particles $M\!=\!64, 256, 1024$ at the initial configurations specified by Eqs. \eqref{4in_a}-\eqref{4in}, with 
$C \in [0.05, 2.5]$, and a wide range of $R_2$ and $R_4$. In Secs.~\ref{basic}-\ref{periodic} we will analyze how the dynamics depends on $R_2$ and $R_4$ for $M\!=\!256$ and $C\!=\!1.5$. Then, in Sec. \ref{otherCN}, we will argue that these results are generic also for the other values of $M$ and $C$.

\subsection{Basic features of the dynamics}\label{basic}
The dynamics of four rings of particles is more complex than the dynamics of two rings. 
Although the general pattern of oscillations combined with settling is kept, in case of four rings the motion in general is not periodic and we 
observe destabilisation and decay of the system, as in 
movie \ref{fig:schem_4rings}, see \cite{movies}. 
Majority of initial conditions lead to the system decay during first $1000$ units of the simulation time, what corresponds to about $100$ oscillations. \par

If the group breaks up, usually one ring is left behind the other three or one ring falls faster than the rest of 
the group; we call it `3+1' type of decay. If the system separates into two pairs of rings, 
we denote it '2+2' type of decay. We observe
also '2+1+1' type of decay when two rings oscillate together and the others are separated. Examples of each decay type 
are presented in 
figure \ref{fig:tra_4}$(a)$-$(c)$ and movies  \ref{fig:tra_4}a-c in \cite{movies}.\par 

For certain initial configurations, 
the particles perform quasiperiodic motion over a very long time.  
It has been checked that 
for several thousands 
of values of $R_2$ and $R_4$ the cluster lifetime exceeds $100 000$.  A typical shape of the corresponding trajectory $Z_i(\rho_i)$ is shown in figure \ref{fig:tra_4}$(d)$ and movie \ref{fig:tra_4}d, see \cite{movies}. These findings 
may indicate that for certain values of $R_2$ and $R_4$, periodic solutions exist, and they will be searched for later in this paper.
\par

We would like to remind that all simulations are performed for symmetrized configurations and the particles by definition do not change the vertical planes they belong to, as described in Sec. \ref{sec:model}. The symmetrisation is done on purpose to find periodic solutions and ensure that the system will not break up because of non symmetrical numerical perturbations before a period is completed. 

\subsection{Decay of the cluster}\label{decay_def}
\begin{figure*}
\includegraphics[width=\textwidth]{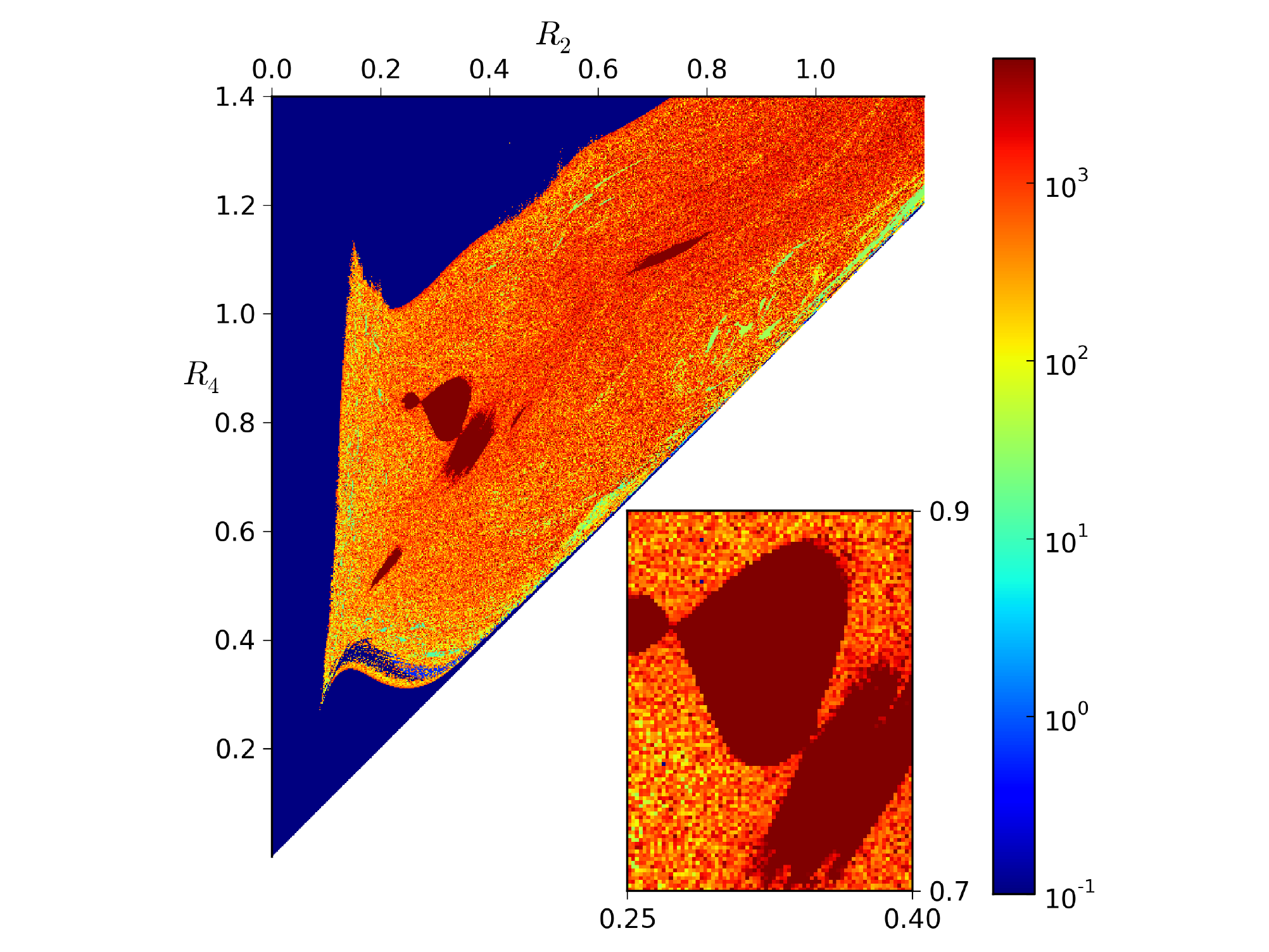}
% \vspace{-0.5cm}
  \caption{(Color) The cluster lifetime for around $420~000$ initial configurations with $C\!=\!1.5$ and different values of $R_2$ and $R_4$ (drawn to scale). The darkest red colour corresponds to the cluster lifetime larger than 
  the simulation time $t\!=\!5000.$ Resolution of the map is $0.002 \times 0.002$ and $M\!=\!256$.}\label{fig:map_lifetime} 
\end{figure*}

\begin{figure*}
\includegraphics[width=\textwidth]{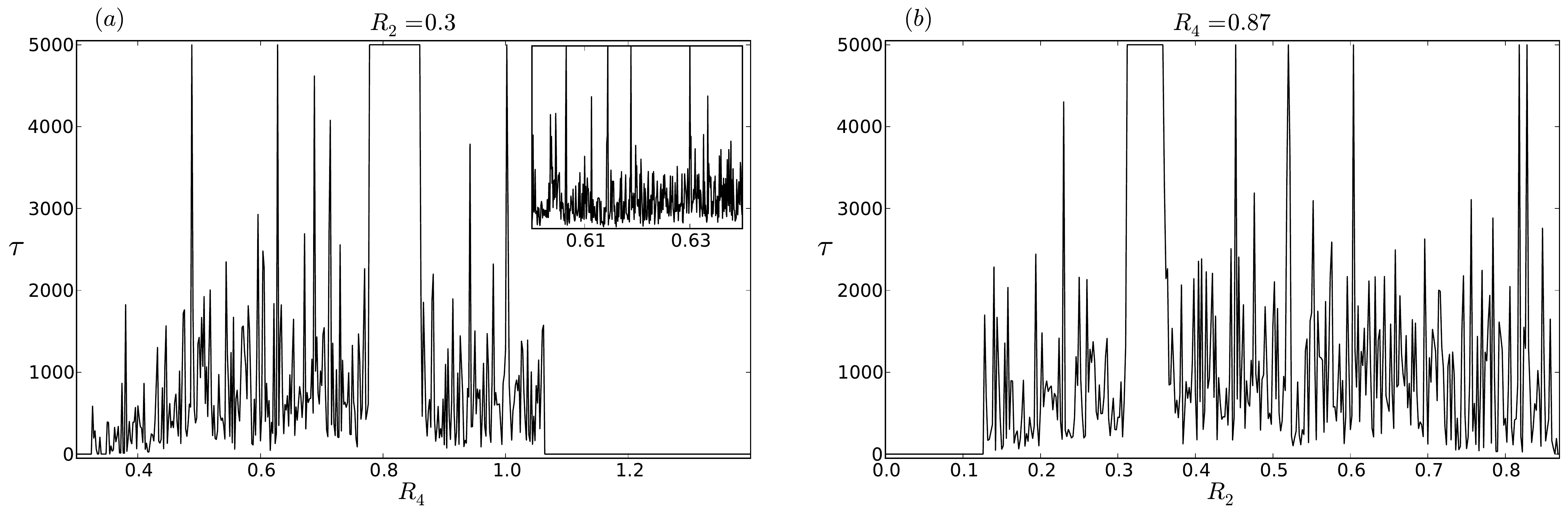}
\caption{Cross-sections (now in the linear scale) through the map of the cluster lifetime from figure \ref{fig:map_lifetime}: (a) $R_2\! =\! 0.3$, (b) $R_4\! = \!0.87$.
 Sensitivity to initial conditions is clearly visible. 
Inset: Scaling down the resolution of $R_2$ and $R_4$ from $0.002$ to $10^{-4}$, a self-similar structure is found.
} \label{fig:crosssectionr2}
\end{figure*}

To describe the system dynamics, it is crucial to define the event of the cluster decay and its time, denoted as $\tau$. 
The intrinsic property of a cluster of particles settling under gravity is the existence of periodic relative motions which may be used as an indicator whether the
particles move together and interact with each other. For this reason we introduce a criterion of a cluster decay based on presence or absence of oscillations between pairs of particles.
In particular, oscillations of particles $i$ and $j$ which belong to the same group imply that the difference of their $z$-coordinates, 
$\Delta z_{ij} = z_{i} - z_{j}$, 
oscillates around zero. We recognise that the 
particles $i,j$ interact and stay together if $\Delta z_{ij}$ has repeating roots.

Therefore, we use the following definition of the cluster decay. 
For each pair of particles $i,j$ we calculate the difference 
$\Delta z_{ij}$ of their $z$-coordinates as a function of time. If for any $l,m$ the time interval between two consecutive roots, $t_A$ and  $t_B>t_A$, of $\Delta z_{lm}$ 
exceeds $1000$, we classify it as the cluster decay and denote $\min_{l,m}t_A$ as the cluster lifetime $\tau$.

This criterion of decay works well, in contrast to attempts based on measuring only the relative vertical distances between rings, because it is common that rings approach each other again, even though they were separated by a very long vertical distance meanwhile, as observed by \cite{EJ} for systems of three particles. 

To define the type of the cluster decay  we apply the following procedure: we count the number of roots of $\Delta z_{ij}$ in the time interval
$[\tau, \mbox{min}(\tau + 1000, T)]$ for each pair of particles $ij$. 
If the number of roots is smaller than the length of the time interval divided by $200$ (for the interval $[\tau, \tau + 1000]$ it is equal to 
$5$), we define that the particles $i,j$ interact with each other. 
Otherwise, we denote them as separated. 
The number $L$ of interacting pairs indicates the type of the decay. $L\!=\!3$ corresponds to `3+1' type of the decay, $L\!=\!2$ to `2+2' type of the
decay, $L\!=\!1$ to `2+1+1' type of the decay and $L\!=\!0$ to `1+1+1+1'. For $L\!=\!6$ there is no decay (type `4').

Now, the main question is how the lifetime depends on initial conditions. The results are presented in figure \ref{fig:map_lifetime} as a map in space of the parameters $R_2$ and $R_4$, with colours indicating the logarithm of the cluster lifetime. 
Values of  
$(R_2, R_4)$ which lead to long-living clusters form a few compact regions, around which only isolated configurations with long lifetimes can 
be found. Deterministic regions are visible when $R_2$ is small enough, or $R_4$ is large enough, and the cluster splits up immediately without oscillations. In the first case, settling velocity of a very small ring 2 is much larger than velocities of the other rings, while 
in the second case, settling velocity of a very large ring 4 is much smaller than velocities of the other rings.

\begin{figure*}
\includegraphics[width=\textwidth]{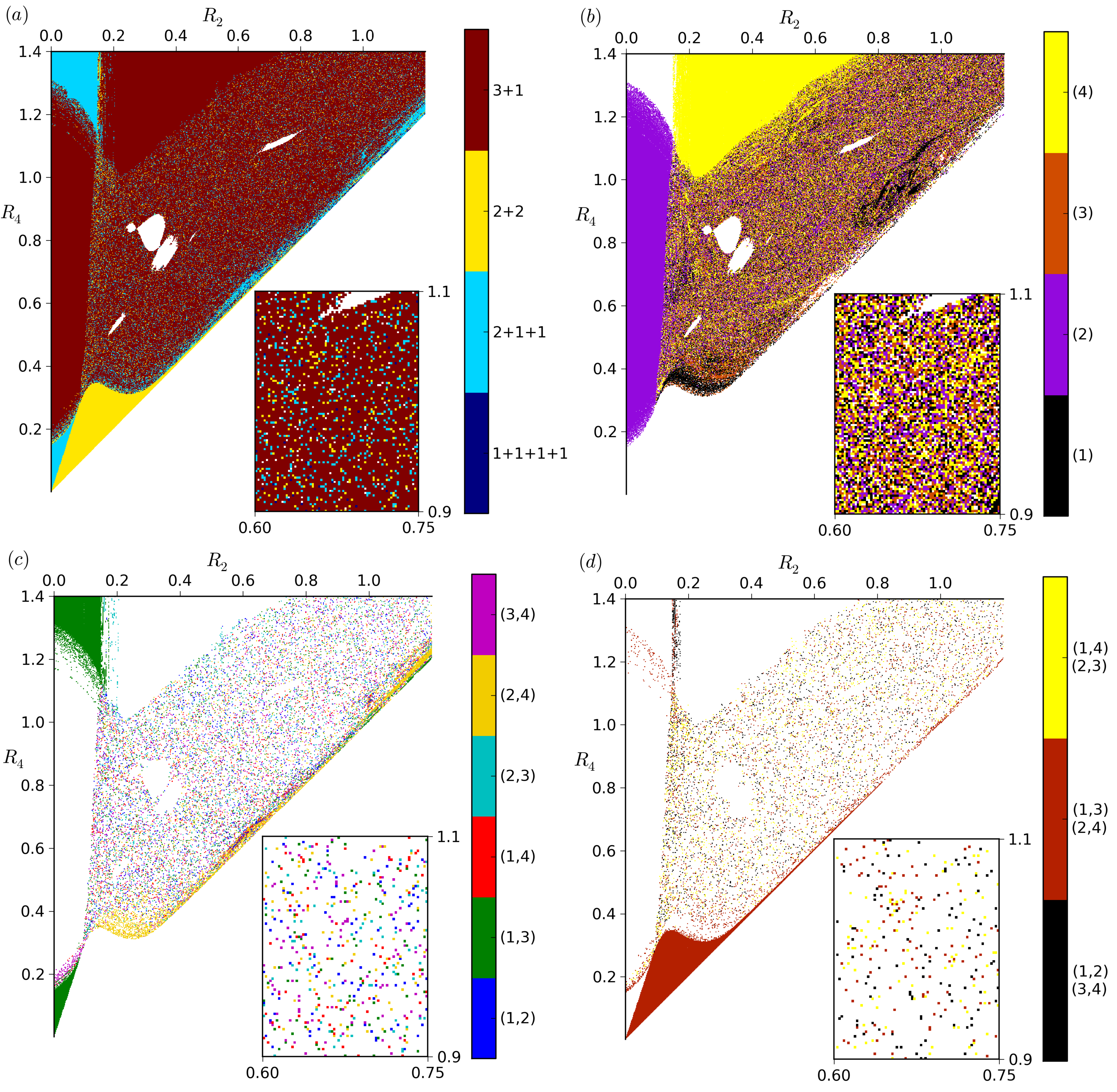}\label{fig:typy_rozpadu_all}
  \caption{(Color) Sensitivity of patterns of the cluster decay to the initial parameters $R_2$ and $R_4$, for  $C\!=\!1.5$ and $M\!=\!256$. White colour means no decay. (a) The types of the decay:  into `3+1', `2+2', `2+1+1' or `1+1+1+1' rings (see Sec. \ref{decay_def} for the exact definitions).    
  (b) `3+1' type of the decay: the colour indicates the label of the ring which separates from the others at time $\tau$. 
(c) `2+1+1' type of the decay: the colours indicate 
the interacting pair of the rings at time $\tau$.
(d) 
`2+2' type of the decay: the colours indicate which rings form interacting pairs at time $\tau$. 
}\label{fig:trajec_4rings_three} 
\end{figure*}

The essential property of the map shown in 
figure \ref{fig:map_lifetime} is that for a wide range of $R_2$ and $R_4$, the cluster lifetime is very sensitive to initial conditions. Around the regions of long-living clusters (with very sharp edges) we find a 
variability of the cluster 
lifetime by orders of magnitude. This property is also well visible in figure  \ref{fig:crosssectionr2} at the cross sections of the map, 
plotted in the linear scale. 
This behaviour is similar to the chaotic scattering reported by \cite{Janosi} for three point particles sedimenting in a vertical plane.

In figure \ref{fig:trajec_4rings_three}, we show 
the type of the cluster decay and labels of the particles which interact with each other at the decay time $\tau$, depending on $R_2$ and $R_4$. 
Sensitivity to initial conditions is clearly visible at most of the map areas, except a few deterministic regions corresponding to a very long or very short lifetimes. 
The last ones can be easily understood by comparing the ring diameters, and taking into account that the smaller the ring, the larger is its velocity.

\subsection{Clusters with long lifetimes and quasiperiodic solutions}\label{periodic}

\begin{figure*}
\includegraphics[width=\textwidth]{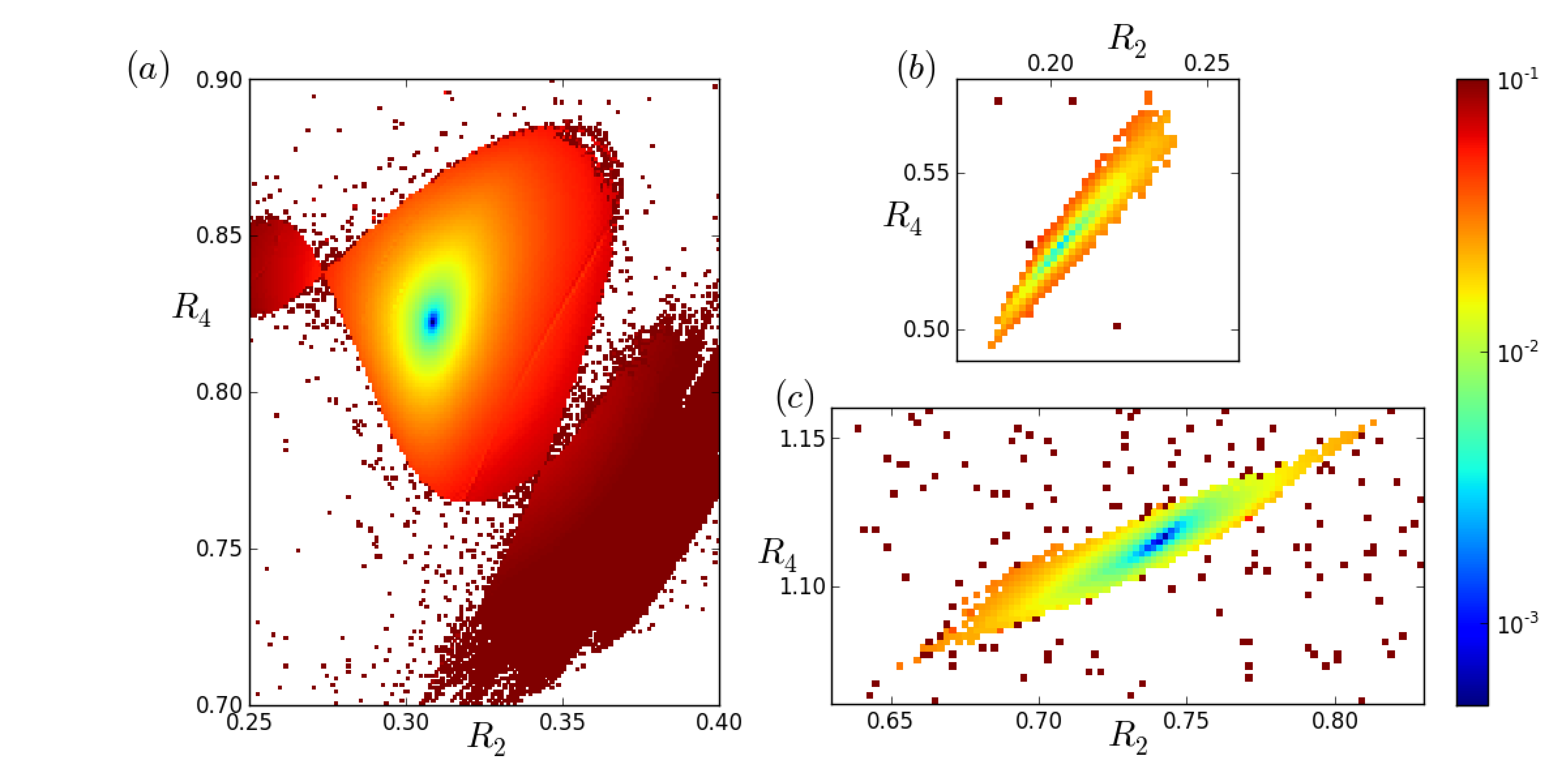}
  \caption{(Color) The average deviation $\Delta$ of particle trajectories for 
 clusters with lifetimes $\tau \ge 5000$,  
$C\!=\!1.5$ and $M\!=\!256$. Resolution: $(a)$: 0.001 $\times$ 0.001; $(b)$ and $(c)$: 0.002 $\times$ 0.002. The minima at the subplots (a),(b),(c) 
correspond to the solutions presented in Fig.8(a),(b),(c), respectively.}\label{fig:trajec_4rings_two} 
\end{figure*} 
We have found that the long lifespan of the cluster and quasiperiodic trajectories of particles are observed for the system with initial configurations grouped in a few regions (see figure 
\ref{fig:map_lifetime}). 
We will now investigate trajectories from each of the specific long-lifetime regions, and  search for periodic solutions.

To this goal, we apply the Dynamic Time Warping (DTW) method, described e. g. by 
\cite{Muller} and \cite{Vlachos}. For each particle $i$, we define its reference trajectory $Z^r_i(\rho_i)$ for times $t_1 \!\le\! t \!\le\! t_3$, where $t_k$, $k\!=\!1,2,3$, are the first, second and third time instants when $Z_i(t_k)\!=\!0$. Then, for $t_3 \!\le\!t\!\le\! 5000$, we calculate the Euclidean distance $d_i(t)$ between the particle position $(Z_i(t),\rho_i(t))$ and its reference trajectory. 
We define the average deviation $\Delta$ of the trajectories as the average of $d_i(t)$ over all the particles $i$ and all times $t$. 

In figure \ref{fig:map_lifetime}, we surround three main long-lifetime islands in the $R_2$-$R_4$ space by rectangular boxes of a fixed resolution, and evaluate for each pixel the average deviation $\Delta$ of the trajectories. 
The results are shown in figure \ref{fig:trajec_4rings_two}. 
For each of the three boxes, there exists a smallest deviation 
$\Delta_{\text{min}}$, corresponding to a very thin trajectory. For example, in figure \ref{fig:trajec_4rings_two}$(a)$,  $\Delta_{\text{min}}\!=\!3 \cdot 10^{-4}$ is three hundred times  smaller than deviation of the quasiperiodic trajectories shown in figure~\ref{fig:tra_4}$(d)$.

\subsection{Three periodic solutions}
\begin{figure*}
\includegraphics[width=13cm]{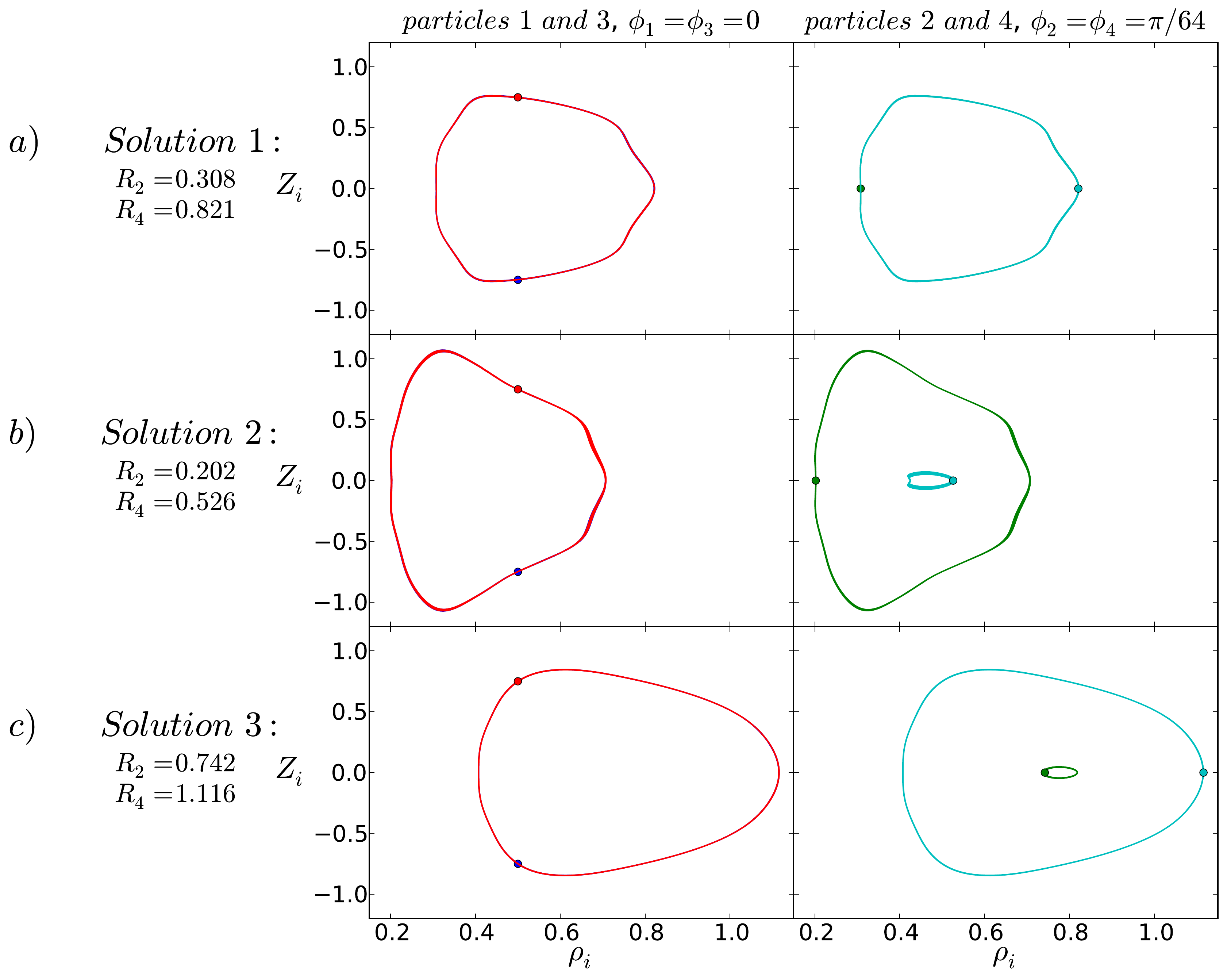}
  \caption{(Color online) Periodic trajectories of four particles (each from a different ring) in the centre of mass  
  frame of a cluster with $M\!=\!256$ particles. Dots: the initial particle positions, with $C\! =\! 1.5$. Solutions in the subplots (a)-(c) correspond to the minima of $\Delta$ in Figs. \ref{fig:trajec_4rings_two}(a)-(c), respectively.
  }
\label{fig:trajec_4rings}
\end{figure*}
The minima visible in figures \ref{fig:trajec_4rings_two}$(a)$,$(b)$,$(c)$ correspond to three periodic trajectories in the centre-of-mass frame of the cluster, shown in figures \ref{fig:trajec_4rings}$(a)$,$(b)$,$(c)$, respectively. 
We remind that owing to the symmetrization of the dynamics, each particle from  
ring $i$ has  
the same coordinates $Z_i(t)$ and $\rho_i(t)$. Therefore, it is sufficient to display $Z_i(\rho_i)$ for particles $i\!=\!1,2,3,4$: each from a different ring $i$, with particles $1,3$ moving in vertical plane $\phi\!=\!0$  
 and particles $2,4$ moving in another vertical plane $\phi\!=\!
4\pi/M$. Here, $M\!=\!256$ and $C\!=\!1.5$.

The new periodic solutions have the following properties.

\textbf{Solution 1}, shown in figure \ref{fig:trajec_4rings}(a) and movie \ref{fig:trajec_4rings}a in \cite{movies}, is located in the middle of the biggest of the long-lifetime regions from figure \ref{fig:map_lifetime}, and at the minimum of the trajectories deviation $\Delta$ in Fig. \ref{fig:trajec_4rings_two}(a). 
All four particles move along the same trajectory $Z_i(\rho_i)$,  shifted in phase by $T/4$, with the period $T=11.7$.

\textbf{Solution 2}, shown in figure \ref{fig:trajec_4rings}(b) and movie \ref{fig:trajec_4rings}b in \cite{movies}, is located in the middle of a small long-lifetime region from figure \ref{fig:map_lifetime}, and at the minimum of the trajectories deviation $\Delta$ in Fig. \ref{fig:trajec_4rings_two}(b). It corresponds to the initial radius $R_2$ 
of the ring 2 
much smaller than  
the initial radius of the rings 1 and 3, $R_1\!=\!R_3\!=\!0.5$, 
and with $R_4 \!\gtrsim \!0.5$. 
The central particle 4 has 
its own tiny trajectory and is circulated by the other three, which 
move along a larger trajectory with the period $T\!=\!10.7$, shifted in phase by $T/3$ with respect to each other. The period of particle 4 is equal to $T/3$.  

\textbf{Solution 3}, shown in figure \ref{fig:trajec_4rings}(c) and movie \ref{fig:trajec_4rings}c in \cite{movies}, 
is located in the middle of the other small long-lifetime region from figure \ref{fig:map_lifetime}, 
and at the minimum of the trajectories deviation $\Delta$ in Fig. \ref{fig:trajec_4rings_two}(c). 
The corresponding initial configuration satisfies the relations,
 %the initial radii 
$R_2,R_4 > R_1\!=\!R_3\!=\!0.5$. 
Now particle 2 is the central one, and 
the motion is qualitatively similar to solution 2, with the interchange of the particles 2 and 4, and much wider rings, what leads to a larger period, $T=15.6$, because larger rings move slower.

\subsection{Two families of periodic solutions}\label{otherCN}

\begin{figure}[h]
\includegraphics[width=5.5cm]{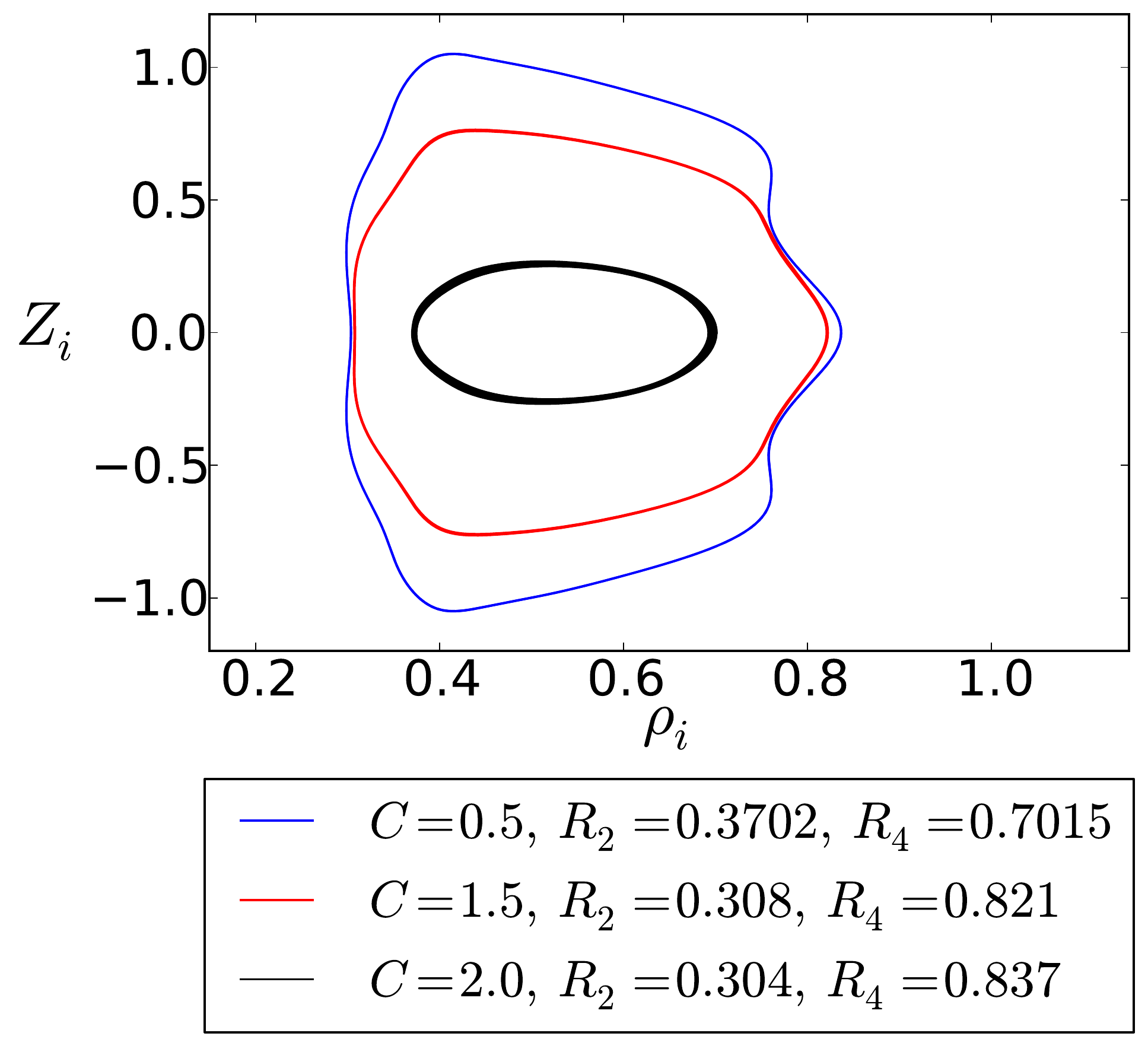}
  \caption{(Color online) Periodic solutions of type 1 
for different values of $C$ 
and $M\!=\!256$.}
\label{fig:fig9}
\end{figure}

Periodic oscillations have been found for many values of $C$. In figure \ref{fig:fig9} we show 
how the shape of the periodic trajectory of type 1 depends on $C$.  Periodic solutions 1, 2 and 3 
for $C\!=\!2$ are shown in figure 10 in the Appendix. These examples illustrate that 
the basic properties of the periodic solutions, described in the previous section and shown in figure \ref{fig:trajec_4rings}, are generic for a wide range of values of $C$. 

Initially, 
two particles are at the same horizontal plane, and the 
other two particles are exactly one above the other. 
Movies \ref{fig:trajec_4rings}b,c in \cite{movies} illustrate that for each periodic solution 2 or 3, 
such a configuration appears again at $T/6$, but with {\it different} values of $C$, $R_2$ and $R_4$, which correspond to solutions 3 or 2, respectively. 
This new configuration can be treated as another 
initial condition 
for the same periodic solution. 
Therefore, assuming that there exist periodic solutions for all sufficiently large values of $C$, 
we predict 
the existence of twin solutions 2 and 3 for each value of $C$ which is large enough.

Dynamics of 4 rings made of other numbers $M\!=\!64,1024$ of the particles have been also evaluated, and similar families of periodic solutions have been found. 

\section{Discussion and concluding remarks}
\begin{figure*}
\includegraphics[width=12.5cm]{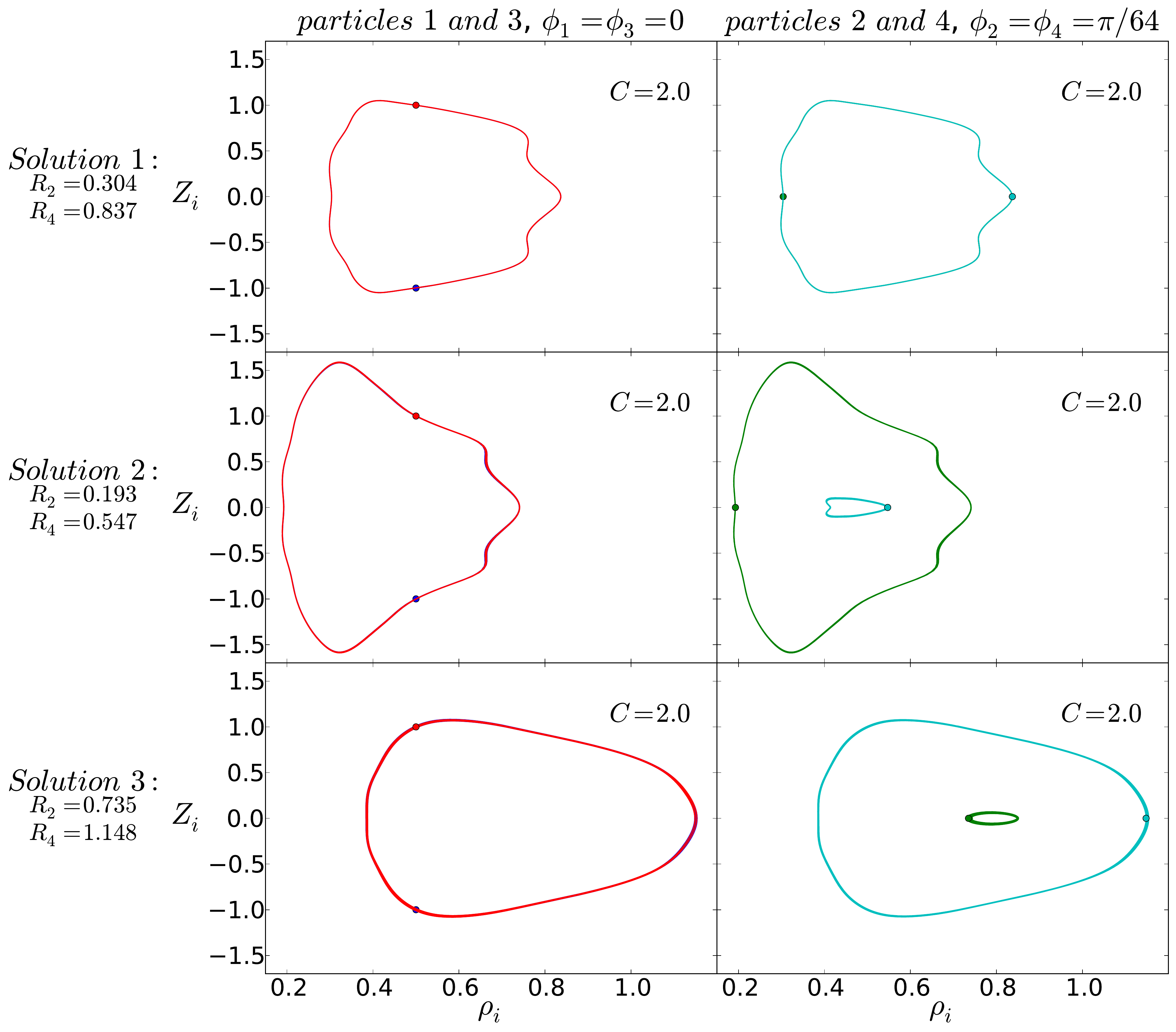}
\caption{(Color online) Periodic trajectories of four particles (each from a different ring) in the centre of mass frame of a cluster with $M\/=\/256$ 
particles. Dots: the initial particles positions. Here, $C\/=\/2.0$ (compare with figure \ref{fig:trajec_4rings} where $C\/=\/1.5$).}
\label{fig:figure8MOD}
\end{figure*}

In this work, we searched for periodic relative motions of many particles which form regular clusters with aspect ratios of the order of one. We generalized in this way the previous solutions found in Ref. \cite{Ekiel-Jezewska} for a smaller number of particles. In Ref.~\cite{Ekiel-Jezewska}, the systems of 2 rings 
were shown to destabilize. Therefore, to determine accurately periodic motions, in this paper we solved the symmetrized dynamics, given in Eqs.~\eqref{EOM}-\eqref{EOM3}. Although we have not analyzed here the non-symmetric perturbations, but it is known from 
Ref. \cite{Ekiel-Jezewska} that such perturbations destabilize the system after times which are large enough to observe the existence of periodic solutions and determine the cluster lifetimes (e. g. for 64 particles, periodic motions destablize at times $t$  not shorter than half of the period, $T/2$, while even the motion during $T/4$ would be sufficient to deduce the existence of unstable periodic solutions). 

A similar behaviour was observed experimentally in Ref.~\cite{Nowakowski} for three close particles sedimenting approximately in a vertical plane. Typically, periodic motions have been observed during times comparable to one sixth of the period, $T/6$. Owing to symmetries of the motion, this is sufficient to demonstrate the existence of unstable periodic orbits. Moreover, in these experiments, the same type of periodic orbit was reached again after the destabilization, and this pattern of approaching and leaving the periodic orbit was repeated several times. 
Summarizing, examples of unstable periodic motions can be easily observed in experiments, and sometimes unstable periodic motions form a generic feature of the dynamics.

In this work, we found a surprisingly large number of periodic solutions. 
For a cluster made of 2 rings, they were observed 
for all the investigated shapes (parameterized by moderate values of the initial aspect ratio $C$),
and particle numbers $N$ even as large as 20 000. 

It has been interesting to modify the initial 2-rings configuration, 
with particles arranged in 4 rings, and in this way desynchronized. 
It turned out that
clusters made of 4 rings usually break up, and their lifetime and type of decay are in general sensitive to initial conditions. In the 2D space of initial parameters, the deterministic area surrounds the chaotic region. The last one contains 
islands of quasiperiodic 
oscillations which do not destabilize during extremely long simulation times. At the centers of these islands, three different periodic solutions exist, parameterized by the initial cluster height and the number of particles.

The results provide a new perspective for the physical mechanism of decay of particle clusters sedimenting in a viscous fluid, typically with a very wide range of life times. %\\\\ \\\\

\newpage
%\vspace{-0.3cm}
\begin{acknowledgments}

%\vspace{-0.4cm}
This work was supported in part by the Polish National Science Centre, 
Grant No. 2011/01/B/ST3/05691. We benefited from scientific activities
of the COST Action MP1305.%\\\\
\end{acknowledgments} 

%\vspace{-0.4cm}
%\newpage
\appendix
\section{Periodic trajectories for a different value of the parameter $C$}% for other numbers of particles $M$ and other values of $C$}

\vspace{-0.4cm}
In this Appendix, we illustrate that Fig.~\ref{fig:trajec_4rings} with $C=1.5$ provides a generic example of two families of periodic solutions: solution 1 corresponds to the first family, and the solutions 2 and 3 to the second family, 
as discussed in Sec.~\ref{otherCN}. Such a structure of the periodic solutions has been observed for many values of the parameter $C$, which determines the initial geometry of the upper and lower rings. In Fig.~\ref{fig:figure8MOD}, 
we show another example: the periodic orbits for $C=2$ (compare with Fig.~\ref{fig:trajec_4rings}).\\\\

\end{document}